\documentclass[thmsa]{article}
\usepackage{amssymb}

\input{tcilatex}
\input tcilatex

\begin{document}

\author{Babur M. Mirza \\
{\small Department of Mathematics, Quaid-i-Azam University,}\\
{\small \ Islamabad 45320, Pakistan} \and Hamid Saleem \\
{\small Theoretical Physics Group, PINSTECH, Nilore, Islamabad, Pakistan; and%
}\\
{\small COMSATS Institute of Information Technology, Islamabad, Pakistan}}
\title{Gravitomagnetic Effects on Collective Plasma Oscillations in Compact Stars }
\date{May 11, 2005}
\maketitle

\begin{abstract}
The effects of gravitomagnetic force on plasma oscillations are investigated
using the kinetic theory of homogeneous electrically neutral plasma in the
absence of external electric or magnetic field. The random phase assumption
is employed neglecting the thermal motion of the electrons with respect to a
fixed ion background. It is found that the gravitomagnetic force reduces the
characteristic frequency of the plasma thus enhancing the refractive index
of the medium. The estimates for the predicted effects are given for a
typical white dwarf, pulsar, and neutron star.
\end{abstract}

\section{Introduction}

The outer crust of a compact star is formed of metal consisting almost
exclusively of iron and other heavy elements. For a typical white dwarf the
solid $^{56}Fe$ core is surrounded by $Si$, $O$, $C$, and $He$ ions and a
gas of free electrons, whereas in neutron stars (pulsars), central densities
are extremely high (up to $10^{12}kg/cm^{3}$) therefore the equation of
state is relatively fixed. The outermost region of a neutron star is formed
of a solid $^{56}Fe$ (density $\approx 10kg/cm^{3}$) crust forming a
crystalline lattice, surrounded by degenerate electrons. The star is thus
regarded as enveloped in a highly ionized degenerate solid state plasma
consisting of densely packed ions and relatively free electrons\cite{[1]}.
Collectively a plasma behaves as an oscillating system having a
characteristic (resonant) frequency called the plasma frequency which for a
degenerate solid state plasma is well defined throughout the medium. In
particular the refractive index of a plasma medium is determined by the
plasma frequency, which in turn determines the transmission and reflection
of radiation for the plasma \cite{[2]}. On the other hand gravitational
effects play a key role in all processes occurring in a compact star, so
that general relativistic effects are important for an adequate description
of the phenomenon. Usually these effects are not directly observable but are
manifest in an indirect way; for example via interaction with the magnetic
field\cite{[3]}, in the accretion of matter\cite{[4]}, and other material
and radiative processes occurring in vicinity of the star\cite{[5]}. Since
in these situations gravitational field of the star is comparatively weak
(as compared to totally collapsed objects), and magnitude of the angular
velocity of the star lies well below the relativistic limit, the equations
of the general theory of relativity can be expressed in an approximate form,
adequate for a description of the gravitational field of the star. This
approximation due to its formal analogy with classical electromagnetic
theory is called the gravitoelectromagnetic (GEM) approximation. The
linearized geodesic equation contains a velocity dependent gravitomagnetic
(GM) force whose origin lies in general relativistic frame dragging effects.
It has been shown \cite{[6]} that due to the gravitomagnetic force a
coupling between the frequency of a material oscillator, consisting of a
point mass, and the angular frequency of a gravitational source establishes.
The coupling is especially enhanced at the resonant frequency of the
oscillator. It is therefore of theoretical as well as observational interest
to study the possible effects of gravitomagnetic force on collective
oscillations of a solid state plasma in compact star crustal regions where
gravitational force is particularly effective.

The paper is organized as follows. In the next section we explain the GEM
approximation to the general theory of relativity and develop a force law,
analogous to classical electromagnetic theory, based on the geodesic
equation. We then extend the formalism for a point mass oscillator, to
include the collective behavior of \ the plasma in a GEM force field, using
the kinetic theory for the plasma oscillations in section 3. The extension
based on the random phase assumption leads to an analogous shift in the
plasma frequency. Later in section 4 we investigate the effect of the
frequency shift on refractive index of the plasma as a single radially
oscillating fluid. In section 5 we summarize the main results and discuss
the possibility of detection of the predicted shift. Throughout we employ
the gravitational units where $G=1=c$ unless mentioned otherwise.

\section{The Gravitelectromagnetic Approximation }

In General Relativity the gravitational field of a massive object is
described by the metric tensor $g_{\alpha \beta }$. For sufficiently weak
gravitating systems, such as the compact stars, a linearization  the metric
can be adequately assumed, where

\begin{equation}
g_{\alpha \beta }=\eta _{\alpha \beta }+h_{\alpha \beta }
\end{equation}
and where $\eta _{\alpha \beta }$ $=diag(-1,-1,-1,1)$ is the Minkowski
metric tensor and $h_{\alpha \beta }$ is the perturbation to the metric such
that $h_{\alpha \beta }\ll 1$ and $x^{\alpha }$ $\equiv (\mathbf{x}%
^{i},x^{0})=(\mathbf{r,}t)$ are the position coordinates of test particle,
with time $t$ being the affine parameter. Further requiring motion to be
slow and $\Gamma _{\beta \gamma }^{\alpha }$ denoting the Christoffel
symbol, the geodesic equation

\begin{equation}
\frac{d^{2}x^{\alpha }}{ds^{2}}+\Gamma _{\beta \gamma }^{\alpha }\frac{%
dx^{\beta }}{ds}\frac{dx^{\gamma }}{ds}=0;
\end{equation}
has the following analogue in the GEM approximation\cite{[7], [8]}

\begin{equation}
\frac{d^{2}\mathbf{r}}{dt^{2}}=\mathbf{G}+\mathbf{v}\times \mathbf{H}
\end{equation}
where 
\begin{equation}
\mathbf{G=-\nabla }\varphi ,\quad \mathbf{H=\nabla \times }4\mathbf{a}
\end{equation}
and

\begin{equation}
\varphi =-\iiint \frac{\rho }{r}dV,\quad \mathbf{a=}\iiint \frac{\rho 
\mathbf{v}}{r}dV
\end{equation}
$\rho $ being mass density and $V$ is the volume. Remarkably the formal
analogy of the above results with the classical electromagnetism extends to
the a gravitational field equations.

Since in the slow rotation approximation the deformations in mass
distribution in the star due to rotation are negligible, the star can be
regarded as a slowly rotating sphere of homogeneous mass density $M$. Then
the gravitoelectric (GE) force acting on a unit mass in vicinity of the star
is given by the Newtonian gravitational force 
\begin{equation}
\mathbf{G=}-\frac{M}{r^{2}}\hat{\mathbf{r}},
\end{equation}
and $\mathbf{H}$ is given by

\begin{equation}
\mathbf{H}=-\frac{12}{5}MR^{2}(\mathbf{\Omega .r}\frac{\mathbf{r}}{r^{5}}-%
\frac{1}{3}\frac{\mathbf{\Omega }}{r^{3}})
\end{equation}
where $R$ is the radius and $\mathbf{\Omega }$ is the angular velocity of
the gravitational source.

The second part of equation (3) has a non-Newtonian origin and is regarded
to exhibit typically general relativistic effects, such as the
Lense-Thirring dragging of frames. Within the GEM approximation the effect
can be interpreted as `gravitomagnetic current' induced in the vicinity of
the gravitational source due to its rotation. It therefore plays an
important role in testing Einstein's theory of gravitation in the weak field
and slow rotation approximation. Further it has been demonstrated\cite{[9]}
that the form of the GM potential does not depend on the choice of a
particular frame or coordinate system used, thus making it physically
significant for all material as well as radiative processes. However to
measure the effects of this force on a given physical system high accuracy
in experiments is required\cite{[10]}.

\section{Equation of Motion for Collective Plasma Oscillations}

A detailed kinetic theory of plasma oscillations in metals has been
developed using the Fermi-Dirac statistics leading to the quantized theory
of elementary excitations\cite{[11]}. However the main features of \
collective plasma oscillations due to the effects of the gravitomagnetic
field can be discussed within \ the classical treatment. We consider the
phenomenon of radial electron oscillations confined to two dimensions,
whereas the angular velocity vector of the star $\mathbf{\Omega }$ is
inclined at an angle $\chi $ to the plane of oscillations. The components of
acceleration due to gravitoelectric force $\mathbf{G}$ to be of constant
magnitude $g$ and the gravitomagnetic force on the $i$th electron has the
magnitude $\dot{r}_{i}Hsin\chi $. The equations of motion for the radial
oscillations of the $i$th electron can thus be written as

\begin{equation}
\ddot{r}_{i}=-\frac{4\pi e^{2}i}{m}\sum_{j,k^{\prime }}(\frac{1}{k}%
)exp[ik(r_{i}-r_{j})]-g+\dot{r}_{i}Hsin\chi .
\end{equation}
where $e$ is the electronic charge and $m$ the electron mass. The prime over 
$k$ denotes a sum where $k=0$ is excluded. Since acceleration due to
gravitoelectric force is of a constant magnitude, its effect on the electron
oscillations with respect to the fixed background of ions will be
negligible. We therefore redefine $\ddot{r}_{i}$ as $\ddot{r}_{i}-g$ in the
equation of motion (8). For all practical purposes electrons can be
considered as point particles, therefore in an arbitrary region of unit
volume the particle density is given by\cite{[12]}

\begin{equation}
\rho (r)=\sum_{i}\delta (r-r_{i}),
\end{equation}
We decompose the particle density function into Fourier components as

\begin{equation}
\rho _{k}(r_{i})=\int \rho (r)exp[-ikr_{i}]dr=\sum_{i}exp[-ikr_{i}],
\end{equation}
such that

\begin{equation}
\rho (r)=\sum_{i,k}exp[ik(r-r_{i})]
\end{equation}
Here it is understood that $i$ as a multiplier is the complex number
denoting $\sqrt{-1}$ and not the script employed for the summation.
Differentiating $\rho _{k}(r_{i})$ with respect to time we obtain

\begin{equation}
\dot{\rho}_{k}(r_{i})=-i\sum_{i}(k\dot{r}_{i})exp[-ikr_{i}],
\end{equation}
and

\begin{equation}
\ddot{\rho}_{k}(r_{i})=-\sum_{i}[(k\dot{r}_{i})^{2}+ik\ddot{r}%
_{i}]exp[-ikr_{i}].
\end{equation}
A substitution from (8) into (13) gives

\begin{equation}
\ddot{\rho}_{k}(r_{i})=-\sum_{i}\{(k\dot{r}_{i})^{2}+ik(-\frac{4\pi e^{2}i}{m%
}\sum_{i,k^{\prime }}(\frac{1}{k^{\prime }})\rho _{k}exp[ik^{\prime }r_{i}]+%
\dot{r}_{i}Hsin\chi )\}exp[-ikr_{i}]
\end{equation}
which can be simplified to give

\begin{eqnarray}
\ddot{\rho}_{k}(r_{i}) &=&-\sum_{i}(k\dot{r}_{i})^{2}exp[-ikr_{i}]-\frac{%
4\pi e^{2}}{m}\sum_{ijk^{\prime }}(\frac{k}{k^{\prime }})\{exp[i(k^{\prime
}-k)r_{i}]\}exp[-ik^{\prime }r_{j}]  \nonumber \\
&&-i\sum_{i}k\dot{r}_{i}Hsin\chi exp[-ikr_{i}].
\end{eqnarray}
Here the second term on the right hand side can be split into two parts one
for $k^{\prime }=k$ and the other for $k^{\prime }\neq k$:

\begin{equation}
\frac{4\pi ne^{2}}{m}\sum_{i}exp[-ikr_{i}]+\frac{4\pi e^{2}}{m}\sum_{k\neq
k^{\prime }}\frac{k}{k^{\prime }}\sum_{i}exp[-ik^{\prime
}r_{i}]\sum_{j}exp[i(k^{\prime }-k)r_{j}].
\end{equation}
The phase factors, randomly distributed in the complex plane, approximately
cancel each other out. Therefore the sum of such factors is small, and hence
the second term being a product of such summations is negligible. Thus
random phase approximation gives in equation (15)

\begin{eqnarray}
\ddot{\rho}_{k}(r_{i}) &=&-\sum_{i}(k\dot{r}_{i})^{2}exp[-ikr_{i}]-\frac{%
4\pi ne^{2}}{m}\sum_{i}exp[-ikr_{i}]  \nonumber \\
&&-i\sum_{i}k\dot{r}_{i}exp[-ikr_{i}]Hsin\chi .
\end{eqnarray}

Further since in a metal electrons are strongly bound to the ions, their
random thermal motion is negligible, therefore the contributions coming from
the first term containing $k^{2}$ are also negligible. Using equations (10)
and (12) in the second and third term of equation (17) we obtain

\begin{equation}
\ddot{\rho}_{k}(r_{i})=-\frac{4\pi ne^{2}}{m}\rho _{k}+\dot{\rho}%
_{k}Hsin\chi .
\end{equation}

\section{Plasma Frequency Shift and Refractive Index}

To determine the resonant frequency for the Fourier component for electron
density function $\rho _{k}(r_{i})$ we assume real functional dependence on
time $t$ in the form of the solution $A_{k}cos\omega t$ to equation (18),
where $A_{k}$ is the amplitude of the density fluctuation and $\omega $ is
applied external frequency. The substitution leads to the following
condition for resonance

\begin{equation}
\omega _{p}^{2}-\omega ^{2}-\omega Hsin\chi =0.
\end{equation}
Since $H\ll \omega _{p}$, solving (19) with $\omega $ unknown and neglecting
terms involving squares and higher powers of $H/\omega _{p}$, we obtain for
the plasma frequency

\begin{equation}
\omega =\omega _{p}-\frac{H}{2}sin\chi .
\end{equation}

Clearly the shift is dependent upon the rotational frequency and mass of the
star, and also upon the angle of inclination $\chi $. It follows from
expression (7) that at the surface of the star $r=R$, the gravitomagnetic
force has magnitude given by

\begin{equation}
H\equiv \mid \mathbf{H}\mid \simeq \mu \sqrt{1+3\cos ^{2}\chi },
\end{equation}
where $\mu =(4GM/5Rc^{2})\Omega $, $\Omega $ being the magnitude of the
angular velocity vector. Substituting from expression (21) into (20) we
obtain an expression relating plasma frequency $\omega _{p}$ to the angle $%
\chi $:

\begin{equation}
\frac{\omega }{\omega _{p}}\simeq 1-\frac{\mu sin\chi }{2\omega _{p}}\sqrt{%
1+3\cos ^{2}\chi },\quad \mu \ll \omega _{p}.
\end{equation}
Here the requirements $H\ll \omega _{p}$ and $\mu \ll \omega _{p}$ are
generally valid for typical compact stellar sources with $\mu $ ranging from 
$0.1687\times 10^{-3}Hz$ (for a typical white dwarf of mass $1M_{\circledast
}=1.989\times 10^{30}kg$ , radius $7\times 10^{6}m$ and angular frequency$1Hz
$) to $236.2932Hz$ (for a typical neutron star of mass $2M_{\circledast }$,
radius $1\times 10^{4}m$, and angular frequency $1kHz$) with corresponding
plasma frequency ranging approximately $5.65\times 10^{6}Hz$ to $5.65\times
10^{6}Hz$ or above. Denoting $\tilde{\omega}_{p}\equiv \omega _{p}-H/2$,
where $H$ is given by (21), we have in terms of plasma frequency we have the
following expression for the index of refraction $\varepsilon $ of the medium

\begin{equation}
\varepsilon (\chi )=\sqrt{1-(\frac{\tilde{\omega}_{p}}{\omega _{a}})^{2}}=%
\sqrt{1-(\frac{\omega _{p}}{\omega _{a}}-\frac{\mu sin\chi }{2\omega _{a}}%
\sqrt{1+3\cos ^{2}\chi })^{2}},
\end{equation}
where $\omega _{a}$ is the angular frequency of waves (e.g. electromagnetic)
propagating through the plasma; whose transmission or reflection, for a
given compact source, is determined by the usual conditions

\begin{equation}
\omega _{a}\leq \tilde{\omega}_{p},\text{ reflection};\text{ \ \ \ }
\end{equation}

\begin{equation}
\omega _{a}>\tilde{\omega}_{p},\text{ transmission}.
\end{equation}

The shift in plasma frequency in the equatorial plane ($\chi =\pi /2$) is $%
\omega _{p}-\mu /2$ whereas in the plane orthogonal to it ($\chi =0$) the
shift is $\omega _{p}-\mu $. Denoting the frequency shift in the equatorial
plane by $\tilde{\omega}_{p\parallel }$ and by $\tilde{\omega}_{p\perp }$
for the orthogonal plane, we find that the maximum difference in the
frequencies transmitted through the plasma:

\begin{equation}
\mid \tilde{\omega}_{p\perp }-\tilde{\omega}_{p\parallel }\mid _{\max }=%
\frac{\mu }{2}.
\end{equation}

In figures (1) and (2) we give plots for the typical cases of neutron star,
pulsar, and white dwarf, between the plasma frequency shift $\omega /\omega
_{p}$ and the angle $\chi $ (using expression (22)) and between the plasma
refractive index $\varepsilon $ as a function of the angle $\chi $ (using
expression (23)) respectively.

\section{Conclusions}

In view of the above investigations we find that the `gravitomagnetic
oscillations' have a coupling to the plasma oscillations which become
especially effective at the plasma frequency. This is so even if the plasma
co-rotates with the star mainly because of the fact, that the
gravitomagnetic force as well as the electron oscillations are along the
radial direction, as is evident from expression (3) and (8). It is clear
from figure (1) and (2) that the plasma frequency the gravitomagnetic force
produces a reduction in the frequency, thus enhancing the refractive index
of the plasma. Further the dependence of the shift on component of the
angular frequency vector of the gravitational source to the plane of
oscillations of the plasma gives a difference in the refractive index, which
results in changing the allowed range of frequencies transmitted through the
plasma, by a maximum amount $\mu /2$. For a typical white dwarf $\mu $ is
approximately $0.1687\times 10^{-3}Hz$, for a typical pulsar $1.6540Hz$ and
for a neutron star $236.2932Hz$. The predicted difference in the frequency
of radiation emitted by a compact star though still very small, may possibly
be observed, for example in radiation spectra of various compact sources, as
the observations become more refined.


\begin{thebibliography}{99}
\bibitem{[1]}  N. K. Glendenning, \textit{Compact Stars}, (Springer-Verlag,
New York, 1997).

\bibitem{[2]}  A. R. Choudhuri, \textit{The Physics of Fluids and Plasmas}
(Cambridge University Press, New York,1998). Ch. 12.

\bibitem{[3]}  L. Rezzolla, B. J. Ahmedov, and J. C. Miller, \textit{Mon.
Not. R. Astr. Soc., }\textbf{322}, 723 (2001) [astro-ph/0011316v2]; A.
Gupta, A. Mishra. H. Mishra and A. R. Prasanna, \textit{Class. Quantum
Grav., }\textbf{15}, 3131 (1998).

\bibitem{[4]}  S. Nitta, M. Takahashi and A. Tomimatsu, \textit{Phys. Rev. }%
\textbf{D44,} 2295 (1991); K. Els\"{a}sser, \textit{Phys. Rev. }\textbf{D62, 
}044007 (2000); A. R. Prasanna and A. Gupta, \textit{Nuovo Cim. }\textbf{B112%
}, 1089 (1997); B. M. Mirza, Charged Particle Dynamics in the Field of a
Slowly Rotating Compact Star [astro-ph/0408579v1].

\bibitem{[5]}  W. K. Rose, \textit{Advanced Stellar Astrophysics}\textbf{\ }%
(Cambridge University Press, New York, 1998).

\bibitem{[6]}  B. M. Mirza, \textit{Inter. J. Mod. Phys}. \textbf{D13}, 327
(2004).

\bibitem{[7]}  M. L.Ruggiero and A. Tartaglia, \textit{Nuovo Cim.} \textbf{%
B117,} 743 (2002)\textit{.}

\bibitem{[8]}  W. Rindler, \textit{Phys. Lett.} \textbf{A 233}, 25 (1997).

\bibitem{[9]}  I. Ciufolini, \textit{Class. Quantum Grav}. \textbf{11} A73
(1994).

\bibitem{[10]}  R. T. Jantzen and G. Mac Keiser, ed. \textit{The Seventh
Marcel Grossmann Meeting on General Relativity}, Part A,(World Scientific,
New Jersey, 1996), pp. 519-529.

\bibitem{[11]}  D. Bohm and D. Pines, \textit{Phys. Rev.} \textbf{92}, 609,
(1953).

\bibitem{[12]}  J. D. Jackson, \textit{Classical Electrodynamics,} (John
Wiley, New York, 1999). Ch.6 \& 7.

\textbf{Figure Captions:}

\emph{Figure1}: Plots for the shift $\omega /\omega _{p}$ in the plasma
frequency of a compact star atmosphere as a function of the angle of
inclination $\chi $ of the plane of observation to the angular frequency
vector for the case of a white dwarf ($M=1M_{\circledast }=1.989\times
10^{30}kg$, $R=7\times 10^{6}m$, $\Omega =1Hz$, $\omega _{p}=$\negthinspace $%
5.65\times 10^{2}Hz$, a pulsar ($M=1.4M_{\circledast }$, $R=$ $3\times
10^{4}m$, $\Omega =30Hz$, $\omega _{p}=5.65\times 10^{4}Hz$),and a neutron
star ($M=2M_{\circledast }$, $R=$ $1\times 10^{4}m$, $\Omega =1kHz$, $\omega
_{p}=5.65\times 10^{6}Hz$)

\emph{Figure 2}: Plots for the refractive index $\varepsilon $ of a compact
star atmosphere as a function of the angle of inclination $\chi $ of the
plane of observation to the angular frequency vector for the case of a white
dwarf ($M=1M_{\circledast }=1.989\times 10^{30}kg$, $R=7\times 10^{6}m$, $%
\Omega =1Hz$, $\omega _{p}=$\negthinspace $5.65\times 10^{2}Hz$, a pulsar ($%
M=1.4M_{\circledast }$, $R=$ $3\times 10^{4}m$, $\Omega =30Hz$, $\omega
_{p}=5.65\times 10^{4}Hz$),and a neutron star ($M=2M_{\circledast }$, $R=$ $%
1\times 10^{4}m$, $\Omega =1kHz$, $\omega _{p}=5.65\times 10^{6}Hz$) for the
propagation of an electromagnetic signal of angular frequency $10^{3}Hz$, $%
10^{5}Hz$, and $10^{7}Hz$ respectively.
\end{thebibliography}
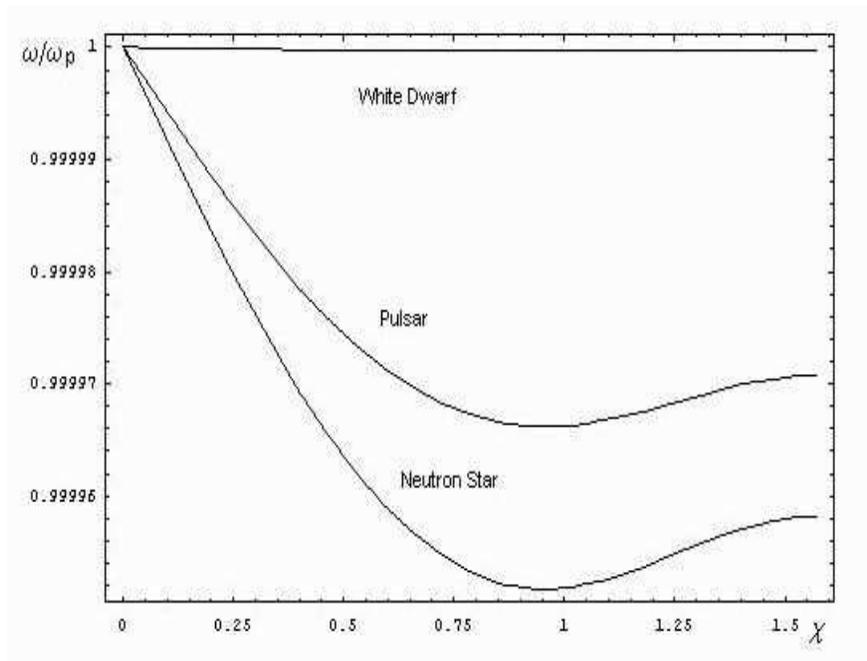
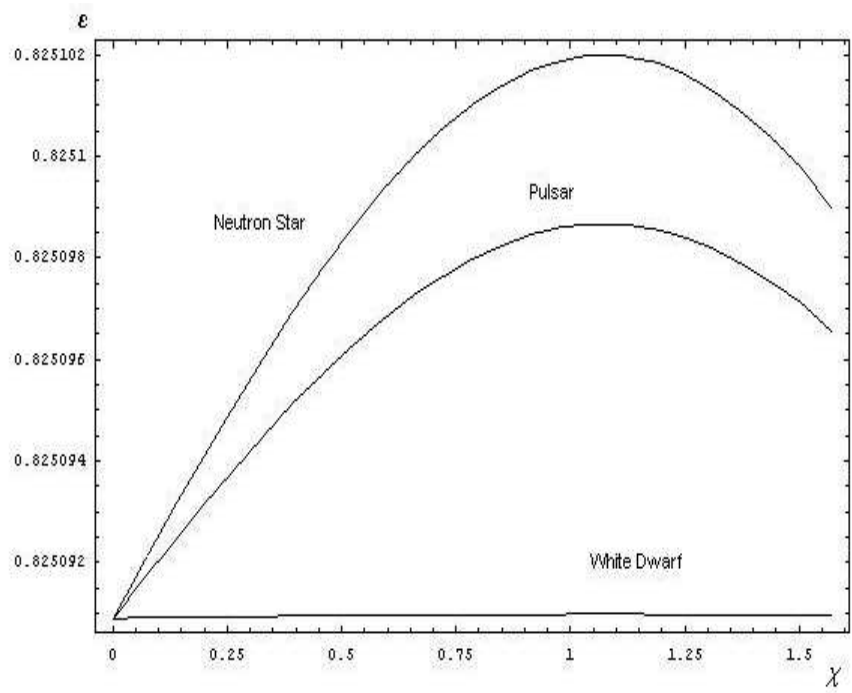
\end{document}